\documentclass[aip,cha,amsmath,amssymb,reprint]{revtex4}

\usepackage{graphicx}
\usepackage{dcolumn}
\usepackage{color}
\usepackage{ulem}
\usepackage{mathptmx, helvet, courier}

\newcommand{\rout}{\bgroup\markoverwith%
{\textcolor{red}{\rule[.5ex]{2pt}{0.5pt}}}\ULon}

\begin{document}

\title{Criteria for folding in structure-based models of proteins}

\author{Karol Wo{\l}ek}
\affiliation{Institute of Physics, Polish Academy of Sciences,
 Al. Lotnik{\'o}w 32/46, 02-668 Warsaw, \\ Poland}

\author{Marek Cieplak}
\email{mc@ifpan.edu.pl}
\affiliation{Institute of Physics, Polish Academy of Sciences,
 Al. Lotnik{\'o}w 32/46, 02-668 Warsaw, \\ Poland}

\date{\today}

\begin{abstract}
\noindent
In structure-based models of proteins, one often assumes that folding
is accomplished when all contacts are established. This assumption
may frequently lead to a conceptual problem that folding takes place
in a temperature region of very low thermodynamic stability, especially when
the contact map used is too sparse. We consider six different structure-based
models and show that allowing for a small, but model-dependent, percentage of the
native contacts not being established boosts the folding temperature substantially
while affecting the time scales of folding only in a minor way. 
We also compare other properties of the six models.
We show that the choice of the description of the backbone stiffness
has a substantial effect on the values of characteristic temperatures that
relate both to equilibrium and kinetic properties.
Models without any backbone stiffness (like the self-organized polymer)  are
found to perform similar to those with the stiffness, including in the studies of stretching.
\end{abstract}



\maketitle

\section{Introduction}

Globular proteins that are found in nature are special sequences
of amino acids that fold rapidly into their native states under
physiological conditions \cite{Anfinsen,ChanDill}. The folding process in
such proteins is considered to take place through motion in a mildly rugged free
energy landscape with a prominent native basin \cite{Wolynes,DillNat,Nymeyer}.
Such a landscape forms through the principles of minimal frustration
and maximal compatibility~\cite{Bryngelson}. 
On the other hand, random sequences are expected to form rough landscapes 
in which many local minima compete for occupancy, as illustrated in a model, 
for instance, in ref. \cite{Garstecki}.

The question: what sequences make good folders, has been
addressed mostly in the context of lattice models and various
criteria have been proposed. One of these
is that the lowest energy state -- the native state -- should have 
large thermodynamic stability~\cite{Shakhnovich}. Another is that
the location of the specific-heat maximum should coincide with that
of the structural susceptibility~\cite{Camacho,Klimov,Folddes}. It 
should be noted that the specific  heat is a measure of fluctuations
in the energy whereas the susceptibility is a measure of fluctuations
in the number of pairs of "amino acids" which stay at their native
spatial separation. A more intuitional criterion has been proposed
by Socci and Onuchic \cite{Socci}. It is formulated in terms of two
temperatures ($T$): $T_f$ and $T_g$. The former is the folding (or melting)
temperature that depends on the energy spectrum and defines a 
temperature below which the probability of occupancy of the
native state, $P_0$, is substantial. The latter relates to the
dynamics and is the $T$ of the glass transition below
which the protein gets trapped  in a non-native state and
folding times, $t_f$, become very long. Bad folders are then
those sequences for which $T_f \;< \; T_g$. The larger the ratio
of $T_f$ to $T_g$, the better the folding properties of the sequence.

The concepts behind this picture appear to be satisfactory in
lattice models where the conformations, including the native one,
are defined in a clear-cut manner and are structurally separated.
Small lattice systems are simple enough to even allow for
Master-equation based  exact analysis of the folding process~\cite{badgo}.
Here, we show that conceptual problems arise 
in off-lattice models and that they can be remedied by slightly relaxing
the criteria for what it means for the system to stay in its native state. 
We focus on the highly studied structure-based, or Go-like, coarse-grained
model~\cite{Go,Nymeyer,Takada} which, by its construction, leads to folding with
a minimal obstruction. This model has various variants 
\cite{Karanicolas,Paci0,Levy,WestPaci,models}
but, invariably, it is defined primarily in terms of the contact
map which specifies which pairs of amino acids form a non-bonded
interaction (most often through a hydrogen bond) in the native state,
which itself may be taken from the Protein Data Bank (PDB)~\cite{pdb}.
The contact maps can be used in all-atom simulations for
descriptive purposes, but in the structure-based models they also
define the dynamics of the system because the contacts are endowed
with effective potentials. The minima of the potentials are located at 
the native separations.

In molecular dynamics studies of folding, one starts from an extended 
conformation and evolves the system until the native state is reached.
Time $t_f$ corresponds to the median first passage time.
In the simplest approach, the native state is declared to be reached
when all of the native contacts become established.
Similarly, $P_0$
is calculated as the probability of all contacts being
established simultaneously in a long equilibrated run and $T_f$
corresponds to a $T$ at which $P_0$ crosses $\frac{1}{2}$. The plot
of $t_f$ as a function of $T$ is typically U-shaped and the $T$
at the center of the U will be denoted as $T_{min}$. The corresponding
optimal $t_f$ will be denoted as $t_{opt}$. We define
$T_g$ operationally as the $T$ at which $t_f\;=\;3\;t_{opt}$ on the
low-$T$ side of the U.

The puzzling observation is that with these simple definitions, $T_{min}$
in off-lattice models
is often found in the region in which $P_0$ is very small -- some
previous examples are in refs. \cite{Hoang2,thermtit,Jaskolski}.
Furthermore, there are many Go-model-based proteins for which the supposedly 
well folding chains are technically bad folders as they have a $T_f$ 
that is smaller or nearly equal to the corresponding value of $T_g$.
Here, we examine  21 proteins within six Go-like models and find that a minor
change in the definition of what it means to be in the native state makes $T_f$ to
coincide with $T_{min}$ in all cases. The change involves requiring that a
fraction, $p$, or more of the native contacts is established simultaneously, 
where $p$, instead of being equal to 1
as in the simple approach, is changed to $0.97 \;\lesssim \;p\;<1 $.
The precise value of $p$ depends on the model and, to a lesser extent,
on the protein. This result indicates that a situation with a few
missing contacts makes the system practically reside within the very
bottom of the native basin. We also find that the small reduction in $p$
leaves the $t_f$s nearly intact while boosting
the apparent thermodynamic stability in a substantial manner.
Relaxing the definition involved in calculating
$P_0$ is expected to enhance the apparent stability. What is surprising,
however, is that making the model system technically a good folder requires
to shift $p$ away from 1 by only 3 percentage points or less.

It should be noted that the concept of the effective $P_0$,
the fraction of time in which all contacts are made
simultaneously,  is quite distinct from
that of the average fraction, $Q$, of the native contacts
that are  present. It is often assumed that the temperature $T_Q$
corresponding to crossing of $Q$ through $\frac{1}{2}$
signals a transition to folding. This assumption
does not seem to have been tested by the actual determination
of the corresponding $t_f$s.  An argument in favor of this idea
can be derived from the behavior of the $Q$-dependent free energy, $F(Q)$,
which develops two minima which are of equal depths at a $T \approx T_Q$.
However, at $Q$ of $\frac{1}{2}$, various subdomains
may be set correctly without a proper establishment of the overall shape 
of the protein -- and hence the relevance of $P_0$ instead of $Q$.

Furthermore, an exact analysis~\cite{transition} of a discretized
$\beta$-hairpin, performed within the so called Munoz-Eaton 
model~\cite{Eaton}, indicates that $Q$ does not constitute a reaction
coordinate even for such a simple system (see also refs.~\cite{badgo,Niewiecz}).
This is related to the fact that Go-like models are defined 
in terms of contacts, but not in terms of $Q$.
It should be noted, however, that all-atom simulations of a 3-helix
bundle interpreted in terms of a Bayesian framework suggest that
using $Q$ is a reasonable approximation in this case~\cite{Bundle}.
In each of the models studied here, $T_Q$ is found to be close
to the location of a maximum in the specific heat. The corresponding 
transition, however, should be associated with an onset of globular conformations
since $T_Q$ is significantly higher than $T_{min}$ and well above the whole
region of $T$s in which folding to one of the globular states -- the
native state -- is appreciable. We find that at $T_Q$
the mean radius of gyration decreases rapidly on cooling.
We also show that the $T$-range of proper
folding and other characteristics, such as mechanostability,
depend on the description of the backbone stiffness.

There are two ways of looking at the folding process. The
perspective taken here is that it is a kinetic phenomenon.
Another is that it is a smoothed out equilibrium phase transition
that should be characterized through trajectories that last much
longer than typical folding times.
Studying $F(Q)$ is an approximate way to describe this transition.
In this paper, we show that the two perspective can be
made compatible if the equilibrium description is focused
on $P_0$, {\it i.e.} on the fraction of time in which nearly
all contacts are present instead of on average number of contacts.
Fluctuations in the total energy do not indicate folding.
There is some analogy to spin glasses here. Phase transitions
in ferromagnets show, in the thermodynamic limit, as a divergence
in the magnetic susceptibility, which is a measure of fluctuations in
magnetization and is related to spin-spin correlation functions. 
However, in spin glasses, it is the nonlinear susceptibility
\cite{Binder,Hertz,Gingras}, and not susceptibility, 
that is sensitive to the transition
from the paramagnetic phase. At the transition, the
susceptibility displays merely a cusp. This nonlinear susceptibility 
is related to different spin correlations.

\section{Structure-based models studied}

The six models we discuss here differ by the selection of the contact
potential and the description of the local backbone stiffness, but the contact
map is common: it is the map denoted by OV+rCSU in ref.~\cite{Wolek},
which is an extension of the OV map. The OV map is derived by considering
overlaps between effective spheres assigned to heavy atoms in the
native state. The radii are equal to the van der Waals radii
multiplied by 1.24~\cite{Tsai}. The rCSU part adds to OV those contacts
which are identified by considerations of a chemical nature. In the case
of the highly mechanostable I27 domain of titin (pdb:1TIT), 
the most important role of rCSU is to add an ionic bridge which
contributes to the strength of a mechanical clamp in this protein.
Generally, OV+rCSU leads to similar and sometimes superior folding 
properties compared to OV. The improvements show
as the broader outline of the U-shaped dependence of
$t_f$ on $T$ and stronger thermal stability. The potentials assigned 
to the contacts between residues $i$ and $j$ are given by
\begin{equation}
V(r_{ij})=\epsilon \; [(\frac{r_{ij}}{\sigma _{ij}})^{12} \;-\;
(\frac{r_{ij}}{\sigma _{ij}})^{\lambda}]
\end{equation}
where $\lambda$ is either 6 (the 12-6 potential) or 10 (the 12-10 potential).
The parameters $\sigma_{ij}$ are derived pair-by-pair from the native
distances between the residues -- the minimum of the potential must coincide
with the $\alpha$-C--$\alpha$-C distance. Consecutive $\alpha$-C  atoms
are tethered by the harmonic potential $k_r\;(r_{i,i+1} -r^n_{i,i+1})^2$,
where $r^n_{i,i+1}$ is the native distance between $i$ and $i+1$. Note that
the stiffness coefficient $k_r$ used here incorporates the customary factor of
$\frac{1}{2}$ in the Hookean energy term. We shall adopt this convention 
in the definition of other harmonic terms discussed below.

We consider three versions of the local backbone stiffness. In many of
our previous works~\cite{JPCM,plos}, we have used the chirality (CH) potential
$V^{CH}=\kappa\varepsilon \;\sum_{i=2}^{N-2}\;(C_i -C_i^{n})^2$
where $N$ is the number of residues and the coeffcient $\kappa$ is set to $\frac{1}{2}$,
as $t_f$ ceases to be $\kappa$-dependent at larger values of $\kappa$~\cite{Kwiatek}.
The chirality of residue $i$ is defined as
$C_i\;=\;(\vec{v}_{i-1}\times \vec{v}_i)\cdot\vec{v}_{i+1}/d_0^3$
where $\vec{v}_i=\vec{r}_{i+1}-\vec{r}_i$ and $d_0$ is the average $r^n_{i,i+1}$.
$C_i^n$ is the native value of $C_i$.

A more common way to account for the stiffness is to combine the bond-angle
term $V_{\theta} = k_{\theta}\;(\theta - \theta _n)^2$ with the dihedral term
$V_{\phi}=K^1_{\phi}[1-cos(\phi-\phi_n)] + K^3_{\phi}[1- cos(3(\phi - \phi_n))]$
where the subscript $n$ indicates the native values. For small departures
of $\phi$ from $\phi_n$, $V_{\phi}=k_{\phi}\;(\phi - \phi_0)^2$,
where $k_{\phi}=\frac{1}{2}(K^1_{\phi}+9K^3_{\phi}$). In this limit, the dihedral
term is equivalent to the chirality potential~\cite{models}.
The values of the coefficients can be derived through all-atom simulations
by matching average conformation-dependent energies to the coarse-grained
expressions. Poma et al.~\cite{Poma} have studied sugar-protein
complexes at room $T$s in this way and, as a byproduct, derived the
following effective values for the parameters pertaining to proteins:  
$\epsilon$=1.5 kcal/mol, $k_r$=100 kcal/(mol {\AA}$^2$), 
$k_{\theta}$=45 kcal/(mol rad$^2$), and  $k_{\phi}$=5 kcal/(mol rad$^2$).
The contacts considered were defined by using the OV approach
and the contact potential was 12-6.
In units of $\epsilon$, these parameters are:
$k_r$=66.66 $\epsilon$/{\AA}$^2$, $k_{\theta}$= 30 $\epsilon$/rad$^2$,
and $k_{\phi}$= 3.33 $\epsilon$/rad$^2$. Assuming, for simplicity,
the equality of the two dihedral coeffcients, we get 
$K^1_{\phi}=K^3_{\phi}$=0.66 $\epsilon$/rad$^2$. This model of
the backbone stiffness will be denoted as AN (for angular).
A similar set of values for the backbone
stiffness has been used by Clementi et al.~\cite{Clementi}
(with a different contact map).  
It is: $k_r$=100 $\epsilon$/{\AA}$^2$, $k_{\theta}$=20 $\epsilon$/rad$^2$,
$K^1_{\phi}$=1 $\epsilon$/rad$^2$, and
$K^3_{\phi}$=0.5 $\epsilon$/rad$^2$.  This model will be denoted as AN'.

Finally, we consider models with no backbone stiffness at all, such as 
the self organized polymer (SOP)~\cite{Hyeon,Mickler}.
In the original SOP model, the harmonic representation of the
peptide bond is expressed as a combination of the 12-6 and FENE
potentials so that they act partially as the 12-6 contact energies
to accelerate the numerics. Here, we use the SOP idea of
not including any backbone stiffness also to the systems
with the 12-10 contacts. We provide comparisons
of protein stretching in the SOP models relative to the models
with finite backbone stiffness. Note that the contact map in the
original SOP model is defined through a cutoff distance whereas we
use OV+rCSU.

The models we study here can be characterized by pairs of symbol
such that the first symbol denotes the contact potential and the
second -- the model of the backbone stiffness. These are:
I -- (12-6, CH),
II -- (12-6, AN),
III -- (12-10, AN),
IV -- (12-10, AN'),
V -- (12-6, SOP),
VI -- (12-10, SOP).
Each of these is studied as described in refs.~\cite{Hoang2,JPCM,plos}
using our own code.
In particular, we use the overdamped Langevin thermostat and the
characteristic time scale in the simulations, $\tau$, is of order 1 ns.
The equations of motion were solved by the 5th order
predictor-corrector method. Due to overdamping, our code is equivalent
to the Brownian dynamics approach.
A contact is considered to be established if its length is 
within 1.5 $\sigma$ or 1.2~$\sigma$ for the 12-6 and 12-10 potentials
respectively~\cite{JPCM,Clementi}.

The equilibrium parameters are determined on at least
3 long trajectories that last for  100 000 $\tau$. 

It should be noted that the all-atom derived coarse-grained parameters
generally depend on the location in the sequence and on the protein
so they should be considered more as an order of magnitude estimates
than precise values.
In particular, the effective $\epsilon$ calculated within
the $\beta$ sheets has been found to be $1.43 \pm 0.3$ kcal/mol and
within the $\alpha$ helices: $1.6 \pm 0.9$~\cite{Poma}. The uniform value
of 1.5 kcal/mol used above has been selected for two reasons:
1) it is characteristic of the turn regions and 2) it coincides
with the strength of a better-binding hydrogen bond.
It corresponds to  755 K so the room
temperature ought to be about 0.4$\epsilon/k_B$. However, due to
the considerable uncertainty in this assessment, our previous
practice has been to assume, when using models with a uniform $\epsilon$,
that the room $T$ situation occurs when folding is optimal as
this is when the model protein behaves as the real one. $\epsilon$ of
1.5 kcal/mol is consistent with 1.55 kcal/mol (or 106$\pm$ 16 pN\;{\AA},
which itself is consistent with 110 pN\;{\AA} of ref. \cite{plos})
that was obtained by matching characteristic values of the force $F_{max}$,
needed to unravel proteins by constant-speed stretching within
the coarse-grained model, to the
experimental results~\cite{Wolek}. This estimate involves making
extrapolations to the experimental speeds and adopting the OV-based contact
map. When using the OV+rCSU map and 38 proteins the effective $\epsilon$
becomes  1.35 kcal/mol (or 93$\pm$14 pN\;{\AA})~\cite{Wolek}.

\section{Results}

\begin{figure}[ht]
\begin{center}
\scalebox{0.4}{\includegraphics{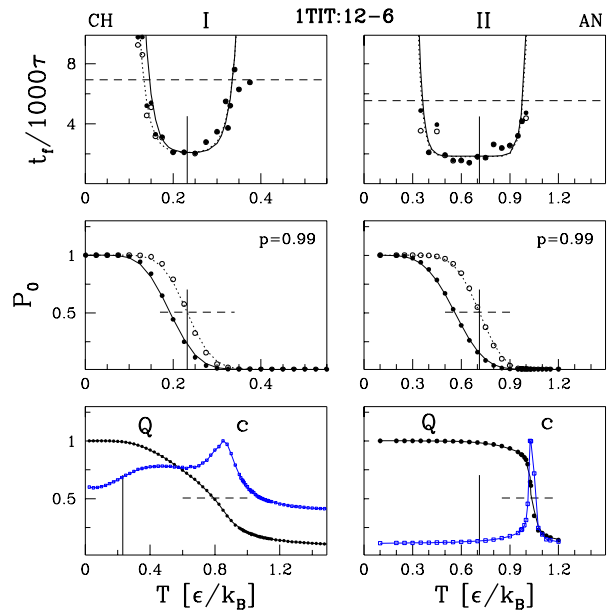}}
\caption{\label{twelve}
Kinetic and thermodynamic properties of models I (the left panels)
and II (the right panels). The explanation of the symbols and lines used
is provided in the main text. In the middle and bottom panels,
the error bars are smaller than the size of the symbols. In the top panels,
a measure of the uncertainty is given by the deviation of the data points
from the fitting lines.}
\end{center}
\end{figure}

\begin{figure}[ht]
\begin{center}
\scalebox{0.4}{\includegraphics{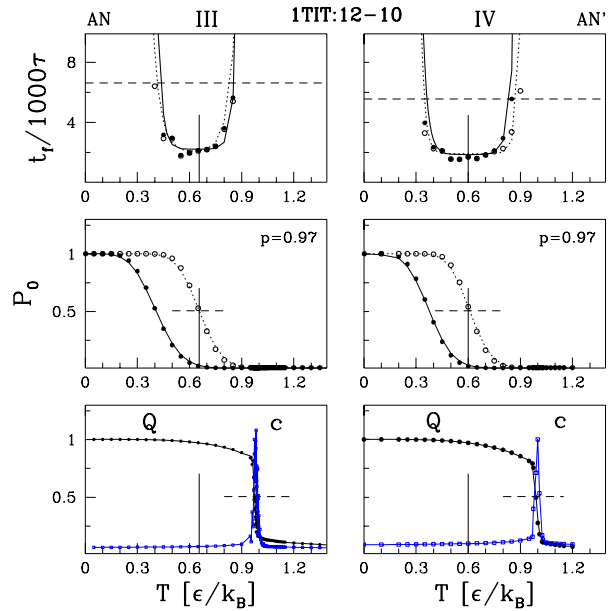}}
\caption{\label{ten}
Similar to Figure \ref{twelve} but for models III and IV.}
\end{center}
\end{figure}

\begin{figure}[ht]
\begin{center}
\scalebox{0.4}{\includegraphics{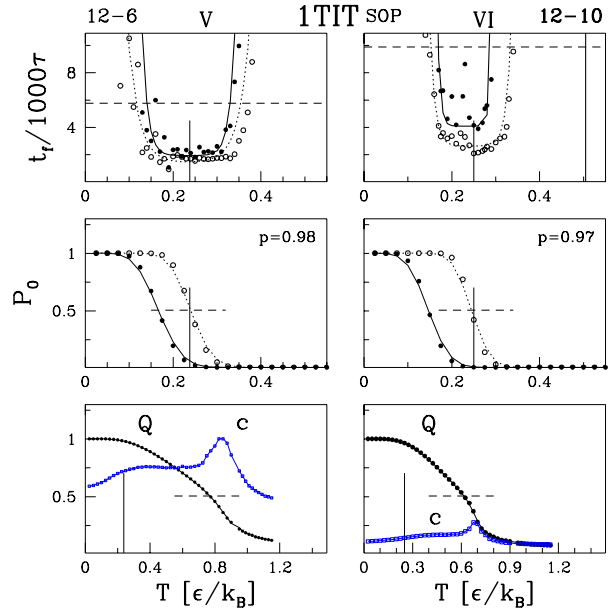}}
\caption{\label{sop}
Similar to Figure \ref{twelve} but for models V and VI.}
\end{center}
\end{figure}

The results pertaining to the thermodynamic and folding properties of the
six models are illustrated in Figures \ref{twelve}, \ref{ten}, and \ref{sop}
for the I27 domain of titin. 
Mechanical stability of this protein has been assessed experimentally in
many studies \cite{Rief,titin,Paci,Oberhauser}.
The OVrCSU contact map for 1TIT consists of 198 contacts.
Table I summarizes  the results for all
of the 21 proteins studied.  The bottom panels show 
the $T$-dependence of $Q$ and $c$ --
the specific heat normalized to its maximal value.
These two quantities do not depend on the choice of $p$. The top
panels show the median $t_f$s and the middle panels show the
effective $P_0$. The general convention is that the solid data
points connected by the solid lines correspond to $p$ of 1
(the undoctored value) whereas the open symbols connected by 
the dotted lines relate to the value of $p$ at which $T_f$ is in the
very middle of the U-shaped folding curve -- it is written in the right-hand
corner of the middle panels.
This value of $T_f$ is indicated by the vertical line. The folding curve
is obtained by fitting the results based on at least 105 trajectories
to the function described by $d_1\cosh(ax+b)+d_2$. The horizontal dashed
lines indicate the level of the optimal $t_f$ multiplied by three from
which an estimate of $T_g$ can be deduced. The horizontal lines in the 
middle and bottom panels indicate the level corresponding to $\frac{1}{2}$
(either for $P_0$ or for $Q$). 
 
The shift in the apparent $T_f$ varies between 0.04 $\epsilon /k_B$ for
model I and 0.25 $\epsilon /k_B$ for model III. It is 0.13, 0.22, 0.06, and
0.11 $\epsilon /k_B$ for models II, IV, V, and VI respectively. The 12-10
potentials are generally effectively more contained, or narrower,
than the 12-6 ones and the effects of $p$ being smaller than 1 are stronger.
Even the shifted effective values of $T_f$ are much lower by at least
50\% than $T_Q$. On the other hand, the separation between $T_Q$ and $T_{max}$
does not exceed 0.07 $\epsilon /k_B$ in any model.
The typical value of $T_{max}$ is of order 1 $\epsilon /k_B$
(though the smallest is 0.7 $\epsilon /k_B$ model VI and then 
it is 0.85 $\epsilon /k_B$ model I) which is puzzlingly high.
The SOP model V has properties which are very similar to that of
model I: the kinetics are close but at $p$=1 the stability of model 
I is higher. We find that when we remove the dihedral term
in model II, while keeping the bond-angle term, $T_{min}$
shifts to a half-way position between models I and II.

\begin{figure}[ht]
\begin{center}
\scalebox{0.4}{\includegraphics{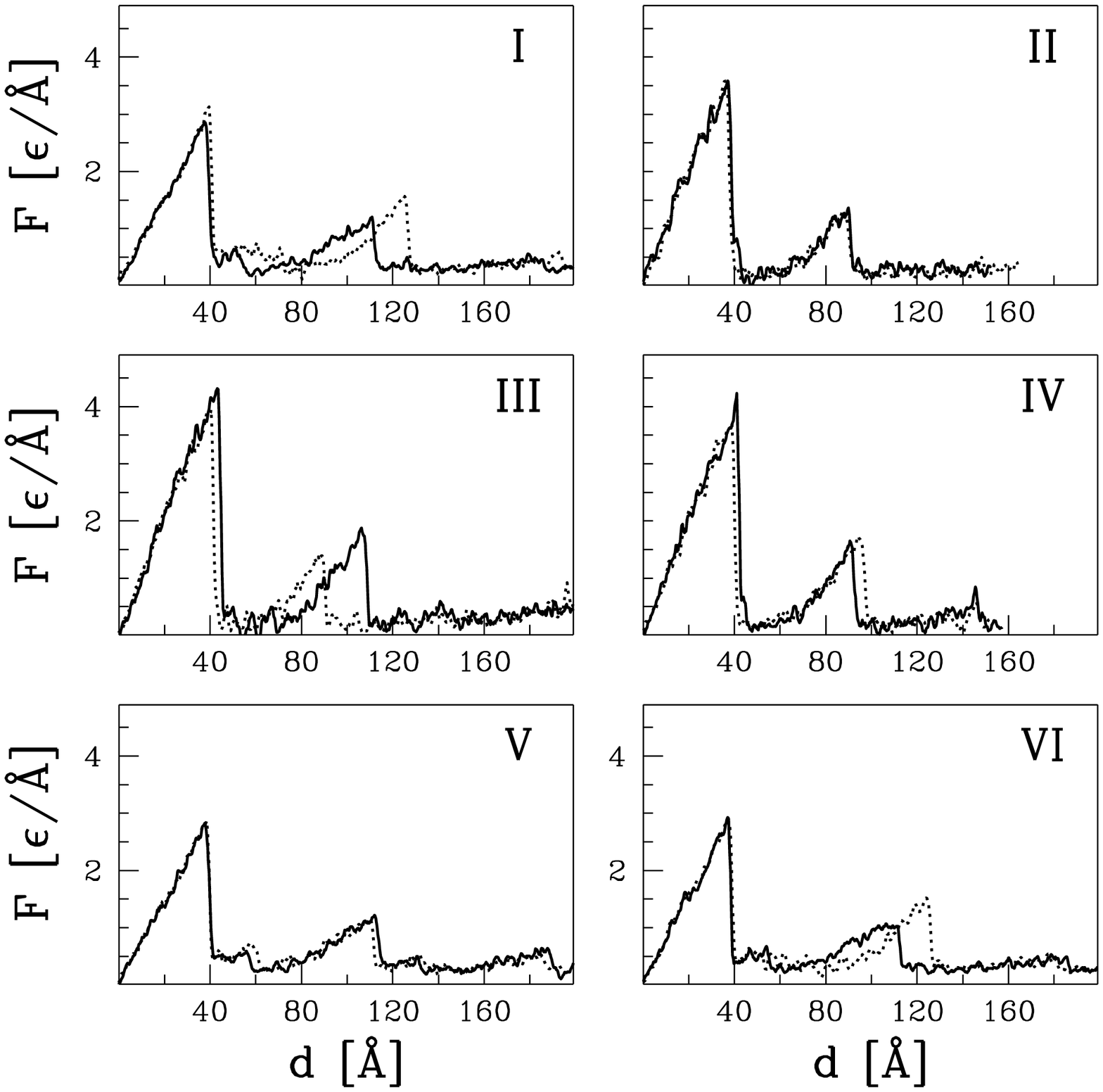}}
\caption{\label{stretch}
Examples of the $F-d$ trajectories for the I27 domain of titin as obtained
by using models I through VI as indicated. In each panel, one trajectory 
at $T$ corresponding to center of the kinetic optimality ($T_{min}$) is 
drawn by a solid line and another by a dotted line. 
The thin (lowest) lines show single trajectories at $T_Q$ -- the force peaks
are at the level of the thermal noise. The values of $T_Q$ are
0.79, 1.04, 0.97, 0.99, 0.77, and 0.61 $\epsilon /k_B$ for models
I through VI respectively.
If one considers a variant of
model V in which the contact map is based on the cutoff distance of
8 {\AA} then one generates 255 contacts instead of 198. This change results
in a similar looking pattern but with a higher $F_{max}$ --  of about 4~$\epsilon$.
Note that the cutoff-based model will have a different calibration of $\epsilon$.
The error bars in $F_{max}$ are of order 0.1~$\epsilon$/{\AA}.}
\end{center}
\end{figure}

We now consider the mechanical stability. Figure \ref{stretch} illustrates
stretching of 1TIT for $T=T_{min}$ and $T=T_Q$.
The terminal residues of the protein
are attached to two harmonic springs~\cite{JPCM,plos}. One
of the springs is anchored at one end and another
is moving at a speed of 0.005~{\AA}/$\tau$.
We monitor the force, $F$, exerted by the protein on the moving
spring as a function of the moving end displacement, $d$. In the initial
state, the protein is set in its native state.
When pulling is implemented at $T=T_{min}$, the $F-d$ patterns
obtained by using the six models are seen to be remarkably similar
though the heights of the force peaks are distinct.
On the other hand, at $T_Q$, the force maxima are
within the noise level. Notice that one cannot model experiments
on stretching by simulating the corresponding kinetics at $T_Q$
(or $T_{max}$ which is very close to $T_Q$).

When comparing mechanostability of many proteins, as in ref.\cite{plos},
it is convenient to choose one fixed $T$ that, for most proteins,
would be within the basin of good foldability, though not necessarily
equal to $T_{min}$. For models I, V, and VI such a temperature is 0.3~$\epsilon /k_B$.
For 1TIT at this $T$, we get
$<F_{max}>$ of 2.04$\pm$0.14, 2.05$\pm$0.12, and 2.44$\pm$0.09 $\epsilon$/{\AA}
respectively when extrapolating to the typical experimental speed of
600 nm/s \cite{titin}. These estimates were obtained based on 100 trajectories.
It should be noted, however, that each model has its own calibration
of the value of $\epsilon$~\cite{Wolek}.
On considering 46 cases of the theory-experiment comparison
(in addition to 38 of ref. \cite{Wolek}, also 1QYS~\cite{1QYS},
2 directions of stretching for 1A1M~\cite{1A1M} and 5 experimental pulling speeds
for 1G6P~\citep{1G6P}), we get the following calibrations for $\epsilon$:
93.1$\pm$15.1, 119.7$\pm$28.0 and 92.2$\pm$11.9 pN~{\AA} for models
I, V, and VI respectively. The corresponding values of the
coefficient of determination, $R^2$,
are 0.846, 0.622 and 0.661, indicating that model I correlates
with the experiment much better than the simpler SOP models.
Similar estimates apply
when stretching at the individually determined $T_{min}$.

In summary, the simple criterion of folding -- that all contacts are
established -- may often lead to a perception that the structure-based
model is unphysical because its thermodynamic stability is substantial in a
low-temperature region in which folding is difficult or absent. This may happen especially
when the contact map has too few elements to be appropriate dynamically, as
is the case of, for instance, the CSU-based contact map~\cite{Sobolev} used in
ref.~\cite{Clementi}. One needs to have a sufficient inter-residue connectivity
for a model protein stay three-dimensional on its own~\cite{Maxwell,Thorpe,Gomez}.
There are many definitions of the contact map and none of them is
perfect. The OV+rCSU map selected here is particularly good, as tested
in ref.~\cite{Wolek} but even this one does not remove the unphysicality.
We have argued here, that allowing for a small percentage of the native
contacts not being established boosts the $T_f$ substantially but
affects the time scales of folding only in a minor way 
(with the exception of model VI).
Determination of the effective thermodynamic stability should then
be less strict (which also helps in hiding imperfections in the 
definition of the contact map) for the model to work well.
The simple strict criterion, however, is simpler to use because it
does not require having the knowledge of the exact value of the
parameter $p$. It should be quite adequate when comparing relative
thermodynamic stabilities of proteins.
The good performance of the SOP models testifies to the fact that
the properties of the structure-based models depends primarily on the
contact map, though the particular choices for the backbone stiffness
affect the optimal folding temperatures.

Model polypeptide chains can be used to describe globular proteins if
folding takes place in the temperature range in which getting to the
native state is associated with a substantial probability of staying
in the vicinity of this state. We have introduced parameter $p$ that
defines what should be meant by the effective native vicinity:
the temperature at which $P_0=\frac{1}{2}$ is at the center of
the U-shaped plot of $t_f$ {\it vs.} $T$.
On decreasing $p$ beyond $\sim$ 0.97, this plot becomes
increasingly broadened, optimal times get shorter, but the center
of the optimal folding kinetics approximately stays unchanged. Another
way to define the relevant vicinity of the native state is to determine
a $p$ at which the optimal $t_f$ is shorter by, say, 10\% compared to
the $p$=1 situation. This approach is harder to implement numerically
because determining a precise value of $t_f$ would require generating
a large number of trajectories. However, qualitatively, it would yield
similar results to what we proposed here.

\vspace{0.5cm}

\noindent {\bf Acknowledgments}

\noindent
We appreciate stimulating discussions with A. B. Poma and B. R\'o\.zycki.
The project has been supported by
the National Science Centre,
Grant No. 2014/15/B/ST3/01905 and by the EU Joint Programme in Neurodegenerative
Diseases project (JPND CD FP-688-059) through the National Science Centre 
(2014/15/Z/NZ1/00037) in Poland.

\vspace*{1cm}

\begin{table}[ht]
\begin{tabular}{|c|ccc|ccc|ccc|ccc|ccc|ccc|}
\hline
      & \multicolumn{3}{c|}{I}         & \multicolumn{3}{c|}{II}& \multicolumn{3}{c|}{III}&  \multicolumn{3}{c|}{IV}&  \multicolumn{3}{c|}{V}&  \multicolumn{3}{c|}{VI}\\
      &$p$&$T_f$&${T_{max}}$    &$p$& $T_f$ &${T_{max}}$   &$p$&$T_f$&${T_{max}}$   &$p$&$T_f$&${T_{max}}$   &$p$&$T_f$&${T_{max}}$   &$p$&$T_f$&${T_{max}}$\\
\hline
1AJ3& 1.00 & 0.22 & 0.47& 1.00 & 0.54 & 0.90& 0.98 & 0.65 & 1.02& 0.99 & 0.45 & 0.93& 1.00 & 0.15 & 0.25& 0.98 & 0.27 & 0.43\\
1ANU& 0.99 & 0.27 & 0.95& 0.99 & 0.79 & 1.12& 0.98 & 0.70 & 1.10& 0.97 & 0.69 & 1.10& 0.99 & 0.25 & 0.93& 0.96 & 0.31 & 0.75\\
1AOH& 0.99 & 0.27 & 0.98& 0.99 & 0.75 & 1.15& 0.98 & 0.68 & 1.12& 0.98 & 0.64 & 1.12& 0.98 & 0.26 & 0.97& 0.97 & 0.29 & 0.77\\
1BNR& 0.99 & 0.30 & 0.73& 0.99 & 0.71 & 1.02& 0.97 & 0.69 & 1.00& 0.97 & 0.64 & 1.00& 0.98 & 0.25 & 0.75& 0.96 & 0.31 & 0.62\\
1BV1& 0.99 & 0.24 & 0.70& 0.99 & 0.66 & 1.02& 0.97 & 0.69 & 1.00& 0.98 & 0.57 & 1.00& 0.99 & 0.20 & 0.80& 0.98 & 0.24 & 0.65\\
1C4P& 0.98 & 0.27 & 0.95& 0.98 & 0.72 & 1.08& 0.97 & 0.71 & 1.02& 0.97 & 0.64 & 1.02& 0.99 & 0.22 & 0.95& 0.96 & 0.28 & 0.77\\
1CFC& 1.00 & 0.20 & 0.57& 0.97 & 0.65 & 0.93& 0.96 & 0.68 & 0.95& 0.94 & 0.63 & 0.93& 0.99 & 0.17 & 0.38& 0.98 & 0.24 & 0.45\\
1EMB& 0.99 & 0.29 & 1.00& 0.99 & 0.78 & 1.20& 0.98 & 0.76 & 1.15& 0.98 & 0.67 & 1.05& 0.99 & 0.26 & 1.00& 0.97 & 0.29 & 0.80\\
1ENH& 1.00 & 0.26 & 0.57& 0.99 & 0.61 & 0.98& 0.97 & 0.68 & 1.05& 0.98 & 0.52 & 1.00& 0.99 & 0.18 & 0.32& 0.97 & 0.28 & 0.45\\
1G1K& 0.99 & 0.29 & 0.98& 0.99 & 0.73 & 1.15& 0.98 & 0.68 & 1.12& 0.98 & 0.63 & 1.15& 0.99 & 0.26 & 0.98& 0.97 & 0.29 & 0.75\\
1OWW& 0.99 & 0.25 & 0.90& 0.99 & 0.72 & 1.08& 0.97 & 0.71 & 1.00& 0.95 & 0.66 & 1.02& 0.99 & 0.24 & 0.90& 0.97 & 0.25 & 0.73\\
1PGA& 0.99 & 0.26 & 0.52& 0.98 & 0.71 & 0.98& 0.97 & 0.70 & 0.98& 0.97 & 0.61 & 0.98& 0.98 & 0.21 & 0.75& 0.97 & 0.24 & 0.43\\
1QJO& 0.99 & 0.26 & 0.88& 0.97 & 0.78 & 1.05& 0.96 & 0.74 & 1.00& 0.95 & 0.71 & 1.00& 0.98 & 0.24 & 0.85& 0.98 & 0.24 & 0.68\\
1QYS& 0.99 & 0.27 & 0.80& 1.00 & 0.68 & 1.02& 0.98 & 0.68 & 1.02& 0.98 & 0.62 & 1.00& 0.99 & 0.22 & 0.80& 0.97 & 0.27 & 0.65\\
1RSY& 1.00 & 0.22 & 0.98& 0.99 & 0.72 & 1.12& 0.98 & 0.68 & 1.08& 0.98 & 0.63 & 1.08& 0.99 & 0.23 & 0.97& 0.99 & 0.23 & 0.77\\
1TIT& 0.99 & 0.23 & 0.85& 0.99 & 0.68 & 1.02& 0.98 & 0.63 & 0.98& 0.97 & 0.60 & 1.00& 0.98 & 0.23 & 0.85& 0.97 & 0.24 & 0.68\\
1TTF& 0.99 & 0.20 & 0.77& 0.98 & 0.62 & 0.93& 0.96 & 0.65 & 0.88& 0.97 & 0.59 & 0.93& 0.99 & 0.20 & 0.75& 0.99 & 0.18 & 0.65\\
1UBQ& 0.99 & 0.27 & 0.77& 1.00 & 0.68 & 1.05& 0.98 & 0.65 & 1.02& 0.98 & 0.61 & 1.02& 0.99 & 0.24 & 0.80& 0.98 & 0.27 & 0.68\\
1WM3& 0.99 & 0.28 & 0.80& 1.00 & 0.66 & 1.05& 0.99 & 0.65 & 1.02& 0.98 & 0.59 & 1.02& 0.99 & 0.24 & 0.82& 0.97 & 0.26 & 0.68\\
2P6J& 1.00 & 0.24 & 0.50& 1.00 & 0.58 & 0.88& 0.97 & 0.65 & 0.95& 0.97 & 0.52 & 0.90& 1.00 & 0.17 & 0.30& 0.98 & 0.24 & 0.40\\
2PTL& 0.99 & 0.22 & 0.80& 1.00 & 0.61 & 0.95& 0.97 & 0.62 & 0.93& 0.98 & 0.55 & 0.93& 0.99 & 0.18 & 0.80& 0.98 & 0.20 & 0.68\\
\hline
Average&0.992&0.253&0.784& 0.990&0.685&1.032& 0.974&0.680&1.019& 0.972&0.608&1.009& 0.989&0.219&0.758& 0.974&0.258&0.641\\
STD    &0.005&0.029&0.161& 0.009&0.058&0.084& 0.008&0.034&0.069& 0.012&0.065&0.065& 0.005&0.030&0.211& 0.009&0.034&0.123\\
\hline
\end{tabular}
\caption{Summary of the results obtained for models I-VI. The protein PDB structure
codes are in the first column. In the case of 1CFC and 2P6J, we consider the first of the 
NMR-derived structures. The original structure file of cohesin 1AOH has missing
several side groups and we use the repaired structure as described in ref.~\cite{Chwastyk}.
In the case on 1EMB, we consider the chromophore molecule to be represented by three mutually 
connected effective amino acids (threonine, glycine, and tyrosine) at locations 65-67.
For each of the models, there are three columns: the column with heading $p$ gives 
the values parameter $p$ at which the kinetic optimum coincides with
$T_f$, the next column lists the values of $T_f$ at this value of $p$, the last
column lists the temperatures corresponding to the maximum in the specific heat.
The temperatures are in the units of $\epsilon /k_B$ and the error 
bars in each entry are smaller than 0.01.
The two bottom lines provide the average over the proteins and the
standard deviation -- for a given model.}
\end{table}
\clearpage

\begin{thebibliography}{30}

\bibitem{Anfinsen}
C. B. Anfinsen, E. Haber, M. Sela, and F. H. White,
Proc. Natl. Acad. Sci. USA {\bf 47}, 1309-1314 (1961).

\bibitem{ChanDill}
H. S. Chan and K. A. Dill,
{\sl The protein folding problem},
Phys. Today {\bf 46}, 24-32 (1993).

\bibitem{Wolynes}
P. G. Wolynes, J. N. Onuchic, and D. Thirumalai,
{\bf 267}, 1619 (1995).

\bibitem{DillNat}
K. A. Dill and H. S. Chan,
Nat. Struct. Biol. {\bf 4}, 10 (1997). 

\bibitem{Nymeyer}
H. Nymeyer, A. E. Garcia, and J. N. Onuchic,
Proc. Natl. Acad. Sci. USA {\bf 95}, 5921-5928 (1998).

\bibitem{Bryngelson}
J. D. Bryngelson and P. G. Wolynes,
Proc. Natl. Acad. Sci. USA {\bf 84}, 7524 (1987).

\bibitem{Garstecki}
P. Garstecki, T. X. Hoang, and M. Cieplak,
Phys. Rev. E. {\bf 60}, 3219-3226 (1999).

\bibitem{Shakhnovich}
M. Karplus, A. Sali, and E. Shakhnovich,
Nature {\bf 369}, 248-251 (1994).

\bibitem{Camacho}
C. J. Camacho and D. Thirumalai,
Proc. Natl. Acad. Sci. USA {\bf 90}, 6369-6372 (1993).

\bibitem{Klimov}
D. K. Klimov and D. Thirumalai,
Phys. Rev. Lett. {\bf 76}, 4070-4073 (1996).

\bibitem{Folddes}
M. Cieplak and J. R. Banavar,
Folding and Design {\bf 2}, 235-245 (1997).

\bibitem{Socci}
N. D. Socci and J. N. Onuchic,
J. Chem. Phys. {\bf 101}, 1519-1528 (1994).

\bibitem{badgo}
M. Cieplak and J. R. Banavar,
Phys. Rev. E. Rapid Comm. {\bf 88}, 040702(R) (2013)

\bibitem{Go}
N. Go, 
Annu. Rev. Biophys. Bioeng. {\bf 12}, 183-210 (1983).

\bibitem{Takada} S. Takada,
Proc. Natl. Acad. Sci. USA {\bf 96}, 11698-11700 (1999).

\bibitem{models} J. I. Su{\l}kowska and M. Cieplak,
Biophys. J. {\bf 95}, 3174-3191 (2008).

\bibitem{Karanicolas}
J. Karanicolas and C. L. Brooks III,
Protein Sci. {\bf 11}, 2351-2361 (2002).

\bibitem{Paci0}
E. Paci, M. Vendruscolo, and M. Karplus,
\textit{Biophys. J.} \textbf{2002}, 83, 3032-3038.

\bibitem{Levy}
Y. Levy, P. G. Wolynes, and J. Onuchic,
Proc. Natl. Acad. Sci. USA  {\bf 101}, 511-516 (2004).

\bibitem{WestPaci}
D. K. West, P. D. Olmsted, and E. Paci,
J. Chem. Phys. {\bf 124}, 154909 (2006).

\bibitem{pdb}
H. M. Berman, J. Westbrook, Z. Feng, G. Gilliland, T. N. Bhat, H. Weissig, I. N. Shindyalov, and P. E. Bourne,
Nucleic Acids Res. {\bf 28}, 235-242 (2000); www.rcsb.org.

\bibitem{Hoang2}
M. Cieplak and T. X. Hoang,
Biophys. J. {\bf 84}, 475-488 (2003).

\bibitem{thermtit}
M. Cieplak, T. X. Hoang, and M. O. Robbins,
Proteins: Struct., Funct. Bioinf. {\bf 56}, 285-297 (2004).

\bibitem{Jaskolski}
M. Chwastyk, M. Jask{\'o}lski, and M. Cieplak,
FEBS J. {\ bf 281}, 416-429 (2014).

\bibitem{transition}
I. Chang, M. Cieplak, J. R. Banavar, and A. Maritan,
Prot. Sci. {\bf 13} 2446-2457 (2004)

\bibitem{Eaton}
V. Munoz and W. A. Eaton,
Proc. Natl. Acad. Sci. USA, {\bf 96}, 11311-11316 (1999)

\bibitem{Niewiecz}
S. Niewieczerza{\l}  and M. Cieplak,
J.Phys.: Condens.Matter {\bf 20}, 244134 (2008).

\bibitem{Bundle}
R. B. Best and G. Hummer,
Proc. Natl. Acad. Sci. USA {\bf 102}, 6732-6737. 

\bibitem{Binder}
K. Binder and A. P. Young,
Rev. Mod. Phys. {\bf 58}, 801-976 (1986).

\bibitem{Hertz}
J. A. Hertz, {\it Spin glasses}, Cambridge University Press,
Cambridge, (1991).

\bibitem{Gingras}
M. J. P. Gingras, C. V. Stager, B. D. Gaulin, N. P. Raju, and J. E. Greedan,
J. Appl. Phys. {\bf 79}, 6170-6172 (1996).

\bibitem{Wolek}
K. Wo{\l}ek, \'A. G\'omez-Sicilia, and M. Cieplak,
J. Chem. Phys. {\bf 143}, 243105 (2015).

\bibitem{Tsai}
J. Tsai, R. Taylor, C. Chothia, and M. Gerstein, J. Mol. Biol.
{\bf 290}, 253-266 (1999).

\bibitem{JPCM} J. I. Su{\l}kowska and M. Cieplak,
J. Phys.: Cond. Mat. {\bf 19}, 283201 (2007).

\bibitem{plos} M. Sikora, J. I. Su{\l}kowska, and M. Cieplak,
PLoS Comp. Biol. {\bf 5}, e1000547 (2009).

\bibitem{Kwiatek}
J. I. Kwieci\'nska and M. Cieplak,
J. Phys.: Condens.Matter {\bf 17}, S1565-S1580 (2005).

\bibitem{Poma}
A. B. Poma, M. Chwastyk, and M. Cieplak,
J. Phys. Chem. B {\bf 119}, 12028-12041 (2015). 

\bibitem{Clementi}
C. Clementi, H. Nymeyer, and J. N. Onuchic,
J. Mol. Biol. {\bf 298}, 937-953 (2000).

\bibitem{Hyeon}
C. Hyeon, R. I. Dima, and D. Thirumalai,
Structure {\bf 14}, 1633-1645 (2006)

\bibitem{Mickler}
M. Mickler, R. I. Dima, H. Dietz, C. Hyeon, D. Thirumalai, and M. Rief,
Proc. Natl. Acad. Sci. USA {\bf 104}, 20268-20273 (2007).

\bibitem{Rief}
M. Rief, M. Gautel, F. Oesterhelt, J. M. Fernandez, and H. Gaub,
Science {\bf 276}, 1109-12 (1997).

\bibitem{titin}
M. Carrion-Vazquez, A. F. Oberhauser, T. E. Fisher, P. E. Marszalek, H. Li,  and J. M. Fernandez,
Prog. Biophys. Mol. Biol. {\bf 74}, 63-91 (2000).

\bibitem{Paci}
S. B. Fowler, R. B. Best, J. L. Toca Herrera, T. J. Rutherford, A. Seward,
E. Paci, M. Karplus, and J. Clarke,
J. Mol. Biol. {\bf 322}, 841-9 (2002).

\bibitem{Oberhauser}
A. F. Oberhauser, P. K. Hansma, M. Carrion-Vazquez, and J. M. Fernandez.
Proc. Natl. Acad. Sci. USA, {\bf 98}, 468-472 (2011).

\bibitem{Chwastyk}
M. Chwastyk, A. B. Poma, and M. Cieplak,
Phys. Biol., {\bf 12}, 046002 (2015).

\bibitem{1QYS}
D. Sharma, O. Perisic, Q. Peng, Y. Cao, C Lam, H. Lu, and H. Li,
Proc. Natl. Acad. Sci. USA, {\bf 104}, 9278--9283 (2007).

\bibitem{1A1M}
B. Sorce, S. Sabella, M. Sandal, B. Samori, A. Santino, R. Cingolani, R. Rinaldi, and P. P. Pompa,
ChemPhysChem, {\bf 10}, 1471--1477 (2009).

\bibitem{1G6P}
T. Hoffmann, K. M. Tych, D. J. Brockwell, and L. Dougan,
J. Phys. Chem. B, {\bf 117}, 1819--1826 (2013).

\bibitem{Sobolev}
V. Sobolev, R. C. Wade, G. Vriend, and M. Edelman,
Proteins: Struct., Funct., Bioinf. {\bf 25}, 120 (1996).

\bibitem{Maxwell}
J. C. Maxwell,
The London, Edinburgh, and Dublin Philosophical Magazine and Journal of Sciences,
{\bf 27}, 294-299 (1864).

\bibitem{Thorpe}
D. J. Jacobs and M. F.  Thorpe,
Phys. Rev. Lett. 75: 4051–4 (1995).

\bibitem{Gomez}
\'A. G\'omez-Sicilia, M. Sikora, M. Cieplak, and M. Carri{\'o}n-V\'{a}zquez,
PLoS Comp. Biol. {\bf 11}, e1004541 (2015); Supporting Information.



\end{thebibliography}
\end{document}